\documentclass[mathleft]{an}
\usepackage{graphicx}
\usepackage{times}
\begin{document}

\Pagespan{789}{}
\Yearpublication{2006}%
\Yearsubmission{2005}%
\Month{11}%
\Volume{999}%
\Issue{88}%

\title{Magnetic flares in Active Galactic Nuclei: Modeling the iron
  K$\alpha$-line}

\author{
R.~W. Goosmann\inst{1,2} \fnmsep
  \thanks{\email{goosmann@astro.cas.cz}\newline},
B.~Czerny \inst{3},
M.~Mouchet \inst{2,4},
V.~Karas \inst{1},
M.~Dov{\v c}iak \inst{1},
G.~Ponti \inst{5,6},
A.~R\'o\.za\'nska \inst{3},
\and
A.-M.~Dumont \inst{2}
}

\titlerunning{Magnetic flares in AGN -- Modeling the K$\alpha$-line}

\authorrunning{Goosmann et al.}

\institute{Astronomical Institute of the Academy of Sciences,
  Bo\v{c}n\'{\i}~II~1401, CZ--14131 Prague, Czech Republic
\and
Observatoire de Paris, Section de Meudon, LUTH, 5 place Jules Janssen,
  F--92195 Meudon Cedex, France
\and
Nicolaus Copernicus Astronomical Center, Bartycka 18, 00-716 Warsaw, Poland
\and
Laboratoire Astroparticule et Cosmologie, Universit\'e Denis Diderot, 2 place
  Jussieu, 75251 Paris Cedex 05, France
\and
Dipartimento di Astronomia, Universit\`a di Bologna, Via Ranzani 1, I--40127
  Bologna, Italy
\and
INAF--IASF Bologna, via Gobetti 101, I--40129, Bologna, Italy
}

\received{30 Aug 2006}
\accepted{---}
\publonline{later}

\keywords{accretion, accretion disks; galaxies: active; radiative transfer;
  relativity; X-rays: galaxies}

\abstract{The X-ray spectra of Active Galactic Nuclei (AGN) are complex and
  vary rapidly in time as seen in recent observations. Magnetic flares above
  the accretion disk can account for the extreme variability of AGN. They also
  explain the observed iron K$\alpha$ fluorescence lines. We present radiative
  transfer modeling of the X-ray reflection due to emission from magnetic
  flares close to the marginally stable orbit. The hard X-ray primary
  radiation coming from the flare source illuminates the accretion disk. A
  Compton reflection/reprocessed component coming from the disk surface is
  computed for different emission directions. We assume that the density
  structure remains adjusted to the hydrostatic equilibrium without external
  illumination because the flare duration is only a quarter-orbit. The
  model takes into account the variations of the incident radiation across the
  hot spot underneath the flare source. The integrated spectrum seen by a
  distant observer is computed for flares at different orbital phases close to
  the marginally stable orbit of a Schwarzschild black hole and of a maximally
  rotating Kerr black hole. The calculations include relativistic and Doppler
  corrections of the spectra using a ray tracing technique. We explore the
  practical possibilities to map out the azimuthal irradiation pattern of the
  inner accretion disks and conclude that the next generation of X-ray
  satellites should reveal this structure from iron K$\alpha$ line profiles
  and X-ray lightcurves.}

\maketitle

\section{Introduction}

X-ray spectra of active galactic nuclei (AGN) are dominated by a power-law
shape with an exponential cutoff at several hundred keV. They frequently show
additional features, such as a soft-excess and signs of warm absorption. The
observed iron K$\alpha$ emission line and the Compton hump in many AGN
indicate a mechanism for X-ray reprocessing. It is widely believed that the
power-law component, i.e. the so-called primary radiation, emerges by
Comptonization in a hot plasma above the disk and is partly reprocessed by the
relatively colder disk atmosphere.

The iron K$\alpha$ line is particularly important to investigate the
inner parts of the accretion flow around black holes. The high cosmic
abundance of iron, together with a relatively high fluorescence yield,
give rise to strong K$\alpha$ line emission. The overall line is
actually a superposition of lines from different ionization states. On top of
that, several effects smear the line shape out, e.g Comptonization inside the
upper layers of the accretion disk, or relativistic and Doppler modifications
in the vicinity of the black hole (see Fabian et al. 2000 or Reynolds \& Nowak
2003 for reviews).

Using theoretical modeling, the observed shape of the iron K$\alpha$ line can
be connected to global parameters of the object. The relativistic broadening
constrains the black hole spin, the radial emissivity profile of the disk, and
the inclination of the object. But also the ``local appearance'' of the line,
i.e. the line profile as it would be seen by an observer co-rotating with the
disk, plays a role. This local profile is determined by the energy dissipation
within the hot plasma and by the structure of the disk.

In this paper we present theoretical modeling of the iron K$\alpha$
line within the frame work of the magnetic flare model (see e.g. Collin et
al. 2003). This model assumes that the hot plasma producing the
primary radiation arises from magnetic reconnection events above the
accretion disk. We consider orbiting flare sources close to the last stable
orbit of a Schwarzschild or a maximally rotating Kerr black hole and we
analyze the spectral appearance and the observed lightcurves for different
azimuthal phases of the orbit.

\section{Model}

\begin{figure}
  \centering
  \includegraphics[width=0.40\textwidth]{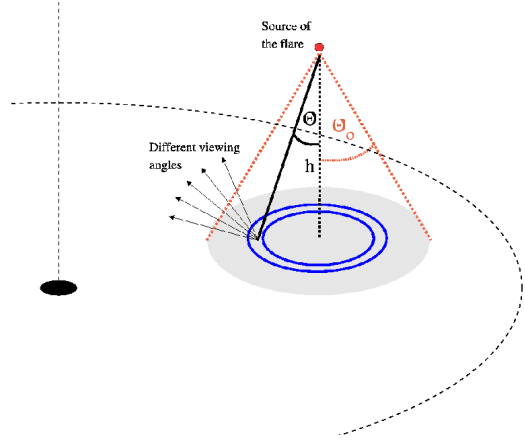}
  \caption{Illustration of the geometrical setup for the flare model. The
  scale of the flare geometry is exaggerated for clarity.}
  \label{fig:flaregeom}
\end{figure}

The flare model for black hole accretion disks was originally
suggested by Galeev, Rosner, \& Vaiana (1979) and then developed in
several papers, such as Haardt, Maraschi, \& Ghisellini (1994). Here we are
specifically interested in the reprocessed component and not in the primary
emission directly seen by a distant observer. This reprocessing has been
extensively modeled in the past and nowadays the models include a detailed
treatment for the hydrostatic equilibrium of the disk (see e.g. Nayakshin,
Kazanas, \& Kallman 2000; Ballantyne, Ross, \& Fabian 2001; R{\' o}{\. z}a{\'
  n}ska et al. 2002). Recently, we added some more sophistication to the
modeling by including the horizontal structure of the irradiated spot
underneath the compact flare source. This is illustrated in
Fig.~\ref{fig:flaregeom}. 

The hot-spot is defined by the height of the flare source above the disk and
by the half-opening angle $\theta_0 = 60\degr$. We divide the spot into 5
concentric rings and for each of the rings we conduct multi-angle radiative
transfer modeling using the codes {\sc Titan} and {\sc Noar} (Dumont et
al. 2000, 2003) in plan-parallel geometry. The primary radiation for each ring
is adjusted for the incident angle and for the illuminating flux as given by
the distance to the point source. A black body component for the accretion
disk is added. We assume that the flare exists for a quarter of an orbit and
evaluate it at four different azimuthal phases. The flare duration is
significantly shorter than the dynamical time-scale of the disk and thus the
hydrostatic equilibrium of the atmosphere should not change over the flare
period. Therefore, the vertical disk profile is computed without external
illumination using a modified version of the code given in R{\' o}{\. z}a{\'
n}ska et al. (2002). The new version includes corrections for the disk
structure due to general relativity. The incident radiation coming from the
flare is by a factor of 144 stronger than the thermal emission from the
accretion disk. We compute locally emitted spectra for a hot spot at $R = 7 \,
{\rm R_g}$, defining ${\rm R_g} = \frac{GM}{c^2}$, from a non-rotating black
hole with $M = 10^8 \,{M_\odot}$. Relativistic and Doppler effects on the
radiation are included by using the ray-tracing code {\sc KY} (Dov{\v c}iak et
al. 2004, Dov{\v c}iak 2004). It takes into account the Doppler velocity due
to the orbital motion of the hot spot and also the intrinsic time evolution of
the emitted light. Here we assume that the primary source is switched on and
off instantaneously. The reprocessed emission evolves from the center of the
spot to the rim.  For further details of the computations see Goosmann 2006 or
Goosmann et al. (in preparation).  

\section{Results}

\begin{figure}
  \centering
  \includegraphics[width=0.40\textwidth]
    {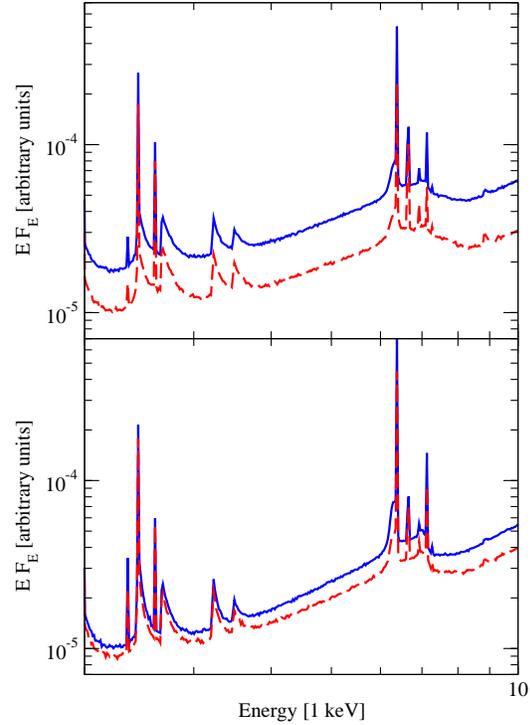}
  \caption{Local reprocessed spectra coming from different locations
  within the illuminated spot at $7 \, {\rm R_g}$. The top panel
  corresponds to the spot center and the bottom panel to the spot
  rim. In each panel the two curves denote local emission angles of
  $\psi = 10\degr$ (blue, solid) and $\psi = 60\degr$ (red, dashed),
  respectively. The emission angle is measured with respect to the disk
  normal.} 
  \label{fig:local-spect}
\end{figure}

Local reprocessed spectra are shown in Fig.~\ref{fig:local-spect} for
the spot center and the rim at two different local emission angles. A
difference in temperature between the two locations in the spot can be
inferred from the spectral slopes and the relative strength of the
components of the K$\alpha$-line. From left to right the ``neutral''
line at 6.4 keV, the helium-like line at 6.7 keV and the hydrogen-like
line at 6.9 keV can be seen. The fourth component represents K$\beta$
line emission. At the spot rim the material is colder due to less
incident flux. The local emission along the disk normal has a harder
spectrum than emission at more grazing angles.

\begin{figure*}
  \centering
  \includegraphics[width=0.44\textwidth]
    {quater-orbit-7rg-combined-H0.5-res800-800_kalpha2.eps}
  \hfill
  \includegraphics[width=0.05\textwidth]{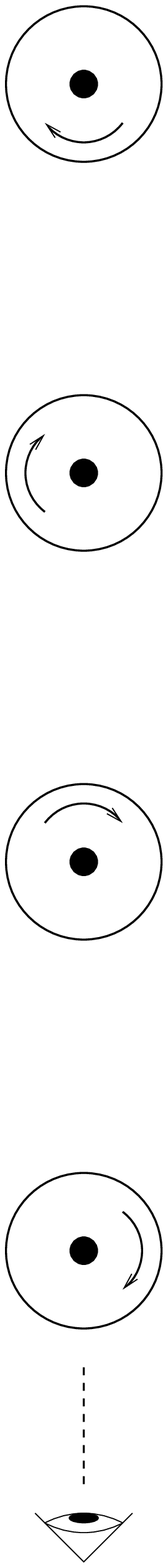}
  \hfill
  \includegraphics[width=0.44\textwidth]
    {quater-orbit-3rg-a0.998-combined-H0.5-res800-800_kalpha2.eps}
  \caption{Integrated reprocessed spectra as seen by a distant observer for a
  hot-spot completing a quarter-orbit close to the marginally stable orbit of a
  black hole. The left part of the figure shows the results for a Schwarzschild
  black hole and a hot spot at $7 \, {\rm R_g}$, the right part for an extreme
  Kerr black hole with $a/M = 0.998$ and a hot spot at $3 \, {\rm R_g}$. The 4
  rows of diagrams represent different azimuthal directions corresponding to
  the central illustrations. As indicated, the three columns of each side
  stand for various inclinations of the distant observer with respect to the
  disk normal.}
  \label{fig:far-spect}
\end{figure*}

The reprocessed spectra seen by a distant observer are plotted in
Fig.~\ref{fig:far-spect} for a Schwarzschild black hole ($a/M = 0$) at a
radial distance of $7 \, {\rm R_g}$ and for the maximally rotating Kerr case
($a/M = 0.998$) at a distance of $3 \, {\rm R_g}$. Note that for the Kerr case
the computations are not entirely consistent as we use again the local spectra
for the Schwarzschild black hole shown in Fig.~\ref{fig:local-spect}. The
differences seen between the two cases plotted in Fig.~\ref{fig:far-spect} are
thus entirely due to the general relativistic and Doppler effects. We applied
the relativistic corrections only to the reprocessed component and neglected
the light-bending acting on the incident radiation during the passage between
the source and the disk. The incident light rays are approximated by straight
lines. This is sufficiently precise for our cases, as the height of the source
above the disk is only $0.5 \, {\rm R_g}$ (see Goosmann 2006 or Goosmann et
al., in preparation).

While both flare locations are close to the marginally stable orbit,
the K$\alpha$ line is more strongly smeared and gravitationally
red-shifted for the Kerr black hole. However, for both cases the individual
line components are hardly recognizable at any azimuthal angle. The orbital
velocity of the spot is higher at $3 \, {\rm R_g}$ than at $7 \, {\rm R_g}$,
which leads to a stronger Doppler shift of the radiation at higher
inclinations. The lensing effect is barely relevant for any of the cases
considered here because the inclination remains moderate ($i \leq 60\degr$),
i.e. the spot is not close to being lined up with the black hole and the
observer.

\begin{figure*}
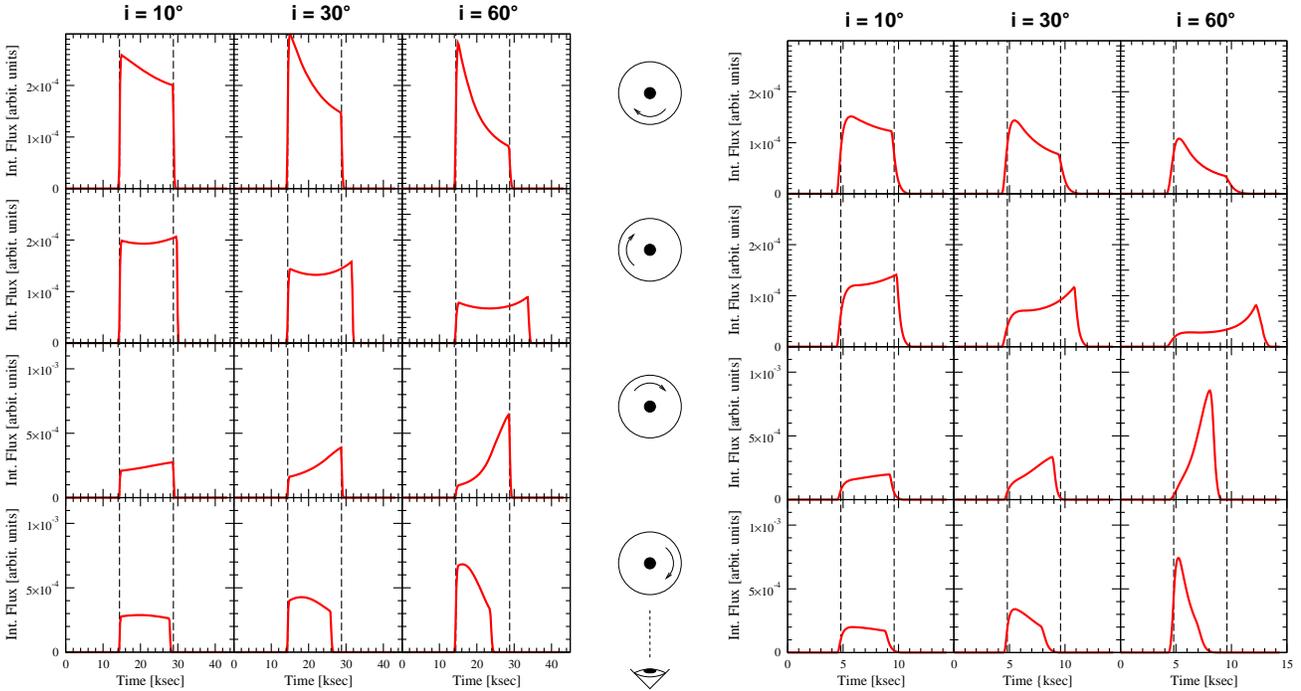

  \centering
  \includegraphics[width=0.44\textwidth]
    {quater-orbit-7rg-combined-H0.5-res800-800_kalpha2-lc.eps}
  \hfill
  \includegraphics[width=0.05\textwidth]{azimuthal-phase-illust.eps}
  \hfill
  \includegraphics[width=0.44\textwidth]
    {quater-orbit-3rg-a0.998-combined-H0.5-res800-800_kalpha2-lc.eps}
  \caption{Lightcurves corresponding to the model spectra shown in
  Fig.~\ref{fig:far-spect}. They were integrated over the energy range of
  2--10 keV. The vertical dashed lines mark the beginning and the end of the
  flare irradiation at the spot center as measured by the distant observer. The
  plots are organized as in Fig.~\ref{fig:far-spect}. Note, that the vertical
  scale differs between the two upper and the two lower rows.}
 \label{fig:far-lc}
\end{figure*}

In Fig.~\ref{fig:far-lc} we present the lightcurves corresponding to the
spectra of Fig.~\ref{fig:far-spect}. The overall shape of the lightcurves is
similar for both rotation states of the black hole, especially at low
inclinations. However, for a given viewing angle the curves differ
significantly for different azimuthal positions of the spot. The timescale for
the evolution of the flare emission across the spot is roughly given by the
height of the source above the disk $H = 0.5 \, {\rm R_g}$. It therefore
corresponds to a light traveling time of $\sim 250 \,$s for $M = 10^8
M_\odot$. The evolution of the local flare emission across the spot is visible
in particular for the rotating black hole. It induces a slight curvature
during the rise and the drop of the lightcurves in Fig.~\ref{fig:far-lc},
while the time evolution of the incident radiation is defined to be
box-shaped. 

\section{Discussion and Conclusion}

We have modeled the spectral shape of the iron K$\alpha$ line and the time
evolution of the X-ray reprocessing for a magnetic flare occurring close to
the marginally stable orbit of a Schwarzschild or a maximally rotating Kerr
black hole. Fig.~\ref{fig:far-spect} and \ref{fig:far-lc} show that, in
principle, the azimuthal location of the reprocessing site close to the last
stable orbit can be effectively constrained for both the Schwarzschild and the
extreme Kerr case. At a given inclination there are differences in the
position of the line centroid caused by Doppler and gravitational
shifts. Also, the line profile and the lightcurve are modified by the various
relativistic and Doppler effects.

In our model we assume a flare duration of a quarter-orbit. The orbital
period at a distance $R$ from a spinning black hole with the normalized
spin parameter $a$ is given by (see the review by Vladim\'ir Karas in this
volume):

\begin{equation}
  T_{\rm orbit} = 3.10 \times 10^3 \, \frac{M}{M_8} \left[ \left(
                  \frac{R}{{\rm R_g}} \right)^\frac{3}{2} + a\right] \;
                  [{\rm s}], \nonumber
  \label{eqn:Torbit}
\end{equation}

\noindent with the mass $M$ given in units of $M_8 = 10^8 \, {\rm
M_\odot}$. One derives from equation (\ref{eqn:Torbit}) that our parameter
values correspond to the following observation times: 

\begin{eqnarray}
  T_{\rm quart}^{\rm Schw} & = & 1.44 \times 10^4 \, \frac{M}{M_8} \; [{\rm
  s}] \; {\rm~for} \; R = 7 \; {\rm R_g}, \frac{a}{M} = 0, \nonumber \\
  T_{\rm quart}^{\rm Kerr} & = & 0.48 \times 10^4 \, \frac{M}{M_8} \; [{\rm
  s}] \; {\rm~for} \; R = 3 \; {\rm R_g}, \frac{a}{M} = 0.998. \nonumber
\end{eqnarray}

With current X-ray satellites observing nearby Seyfert galaxies with high
count-rates a sufficiently detailed shape of the iron line requires
observation times of $\sim 20$ ksec or more. Therefore, a flare at 3 ${\rm
R_g}$ lasting for a quarter of an orbit cannot be observed yet. The
observational limit might be obtained for massive black holes and slightly
larger radii though. The next generation of X-ray satellites like {\sc Xeus}
or {\sc Constellation-X} is planned to have much larger effective collecting
areas. These missions should therefore allow to map out the innermost
azimuthal irradiation pattern of some AGN accretion disks.

Meanwhile, variability modeling is another possibility to compare the type of
radiative transfer calculations presented here to observed data. An example
is given in Goosmann et al. (2006), where we model the rms variablity
spectrum of the Seyfert galaxy MCG-6-30-15. The modeling confirms that the
central black hole in this object is rapidly spinning and that the radial
profile of energy generation in the hot corona of the accretion disk must rise
steeply toward the disk center. Further steps in this line of research are
possible when including also the power density spectrum that has become
available for some nearby Seyfert galaxies such as NGC~3516 (Edelson et
al. 1999), NGC~4051 (McHardy et al. 2004), or MCG-6-30-15 (McHardy et
al. 2005).

\acknowledgements
This research was supported by the Center for Theoretical Astrophysics in
Prague and by the Polish research grant 1P03D00829. RWG also thanks the
Hans-B\"ockler-Stiftung.

\end{document}